\documentclass[twocolumn,showpacs,showkeys,superscriptaddress]{revtex4}

\usepackage{amsmath}
\usepackage{amssymb}
\usepackage{graphicx}
\usepackage{color}
\usepackage{verbatim}

\begin{document}

{\raggedleft {\it Accepted for publication in Physics of Plasmas}}
\\

\title{Extended theory of the Taylor problem in the plasmoid-unstable regime}

\author{L. Comisso$^{1,*}$, D. Grasso$^{1}$, F.L. Waelbroeck$^2$\\
{\small $^1$ Dipartimento Energia, Politecnico di Torino, Corso Duca degli Abruzzi 24, 10129, Torino, Italy, \\
and Istituto dei Sistemi Complessi - CNR, Via dei Taurini 19, 00185, Roma, Italy} \\
{\small $^2$ Institute for Fusion Studies, The University of Texas at Austin, Austin, TX 78712-1203, USA} \\
{\small $^*$ Electronic mail: luca.comisso@polito.it}}

\begin{abstract}
A fundamental problem of forced magnetic reconnection has been solved taking into account the plasmoid instability of thin reconnecting current sheets. In this problem, the reconnection is driven by a small amplitude boundary perturbation in a tearing-stable slab plasma equilibrium. It is shown that the evolution of the magnetic reconnection process depends on the external source perturbation and the microscopic plasma parameters. Small perturbations lead to a slow nonlinear Rutherford evolution, whereas larger perturbations can lead to either a stable Sweet-Parker-like phase or a plasmoid phase. An expression for the threshold perturbation amplitude required to trigger the plasmoid phase is derived, as well as an analytical expression for the reconnection rate in the plasmoid-dominated regime. Visco-resistive magnetohydrodynamic simulations complement the analytical calculations. The plasmoid formation plays a crucial role in allowing fast reconnection in a magnetohydrodynamical plasma, and the presented results suggest that it may occur and have profound consequences even if the plasma is tearing-stable.
\end{abstract}

\pacs{52.35.Vd, 52.30.Cv, 52.35.Py}

\keywords{magnetic reconnection, magnetohydrodynamics }

\maketitle

\section{Introduction}

In 1985, Hahm and Kulsrud published a seminal paper \cite{HK_1985} on a basic problem of forced magnetic reconnection, the so-called ``Taylor problem'' (after the physicist J.B. Taylor who first proposed it). In this model of forced reconnection, a tearing-stable slab plasma equilibrium is subject to a small-amplitude boundary perturbation that drives magnetic reconnection.
Adopting a resistive magnetohydrodynamic (MHD) description of the plasma, Hahm and Kulsrud showed that after an initial phase in which a current sheet builds up, the forced reconnection process evolves according to a linear resistive phase and then a nonlinear Rutherford regime \cite{Rutherford_1973}.

Successively, in 1992, Wang and Bhattacharjee \cite{WB_1992} showed that under certain conditions the reconnection process in the Taylor model passes into the nonlinear regime according to a Sweet-Parker-like evolution \cite{Waelbroeck_1989} instead of a Rutherford one. These two scenarios have been successively investigated for the case of a visco-resistive MHD plasma by Fitzpatrick \cite{Fitz_2003}. In particular, he determined an expression for the threshold perturbation amplitude required to trigger a Sweet-Parker-like evolution. 

Since the pioneer work by Hahm and Kulsrud, the Taylor problem has not been free from controversies \cite{Ishizawa_2000,Ishizawa_2001,BFW_2001,IT_2001,Cole_2004,Vekstein_2004,ColeFitz_2004,Fitz_2008} and has continued to be of great interest in plasma physics \cite{WangBhatta_1992,Ma_1996,Vekstein_1998,Avinash_1998,Rem_1998,Bulanov_1998,Vekstein_1999,Valori_2000,FBML_2003,
Fitz_2004a,Fitz_2004b,Bian_2005,Birn_2005,Vekstein_2006,Birn_2007,HossVek2008,Gordo_2010a,Gordo_2010b,LazzCom2011,
Hosseinpour2013,Dewar_2013}. In fact, Taylor's model has considerable implications for laboratory plasmas (such as those in tokamaks, in which resonant magnetic perturbations can lead to magnetic reconnection) as well as astrophysical plasmas (such as those characterizing the solar corona, in which magnetic reconnection can be forced by the motion of the photospheric footpoints). However, despite the numerous works related to this problem, the nonlinear possible scenarios of the magnetohydrodynamical Taylor model have not changed from those found by Hahm and Kulsrud \cite{HK_1985}, Wang and Bhattacharjee \cite{WB_1992}, and Fitzpatrick \cite{Fitz_2003}.

In this article, extending our preliminary results \cite{CGW_2014}, we give evidence that a different nonlinear scenario is possible. We show that under certain conditions, after the initial phase in which a very elongated current sheet forms at the resonant surface, there is a sudden transition to a fast reconnection regime characterized by the disruption of the current sheet due to the development of plasmoids. In spite of the fact that it is known from quite a long time that elongated reconnecting current sheets can become unstable to the formation of plasmoids \cite{Matt_Lam_1985, Bisk_1986, Lee_Fu_1986, Fu_Lee_1986}, this possibility has not been taken into account in the theory of forced magnetic reconnection in Taylor's model. In other contexts, instead, there has been a growing interest in the investigation of the link between plasmoid formation and fast reconnection \cite{Daugh_2006,Lou_2007, Daugh_2007,BHYR_2009, SLUSC_2009, Daugh_2009, Cassak_2009, SC_2010, Skender_2010, ULS_2010, HB_2010,HBS_2011,Ni_2012,Ni_2012b, LSU_2013, Murphy_2013,Pucci_2014,Graf_2014,
Yu_2014,Ni_2015,Nemati_2015,Tenerani_2015}. 

Here, we extend previous analysis of the Taylor problem considering also the plasmoid-unstable regime. We discuss all the scenarios of the collisional magnetohydrodynamical Taylor model within a unified framework and we examine the conditions under which these scenarios occur. Both resistivity and viscosity are taken into account, since both may play an important role in laboratory as well as in astrophysical environments \cite{Park_1984,Craig_2012,Tenerani_2015}. We perform both analytical and numerical calculations, showing that the development of plasmoids causes the onset of a fast reconnection regime. Therefore, the present analysis is important to understand the evolution of the collisional reconnection process from the current sheet formation to fast magnetic energy conversion rates.

\section{Basic Equations and Taylor's Problem}  \label{sec2}

We consider a plasma described by the two-dimensional (2D) reduced MHD model \cite{Strauss_1976}. Within this description the magnetic and velocity fields can be expressed in terms of flux and stream functions as
\begin{equation} 
{\bf{B}} (x,y,t) = {B_z}{{\bf{e}}_z} + \nabla \psi (x,y,t) \times {{\bf{e}}_z} \, ,\label{e_B}
\end{equation}
\begin{equation} 
{\bf{v}} (x,y,t) = {{\bf{e}}_z} \times \nabla \phi (x,y,t) \, ,\label{e_V}
\end{equation}
where ${{\bf{e}}_z}$ is the unit vector along the ignorable coordinate. The flux and stream functions are governed by the resistive Ohm's law
\begin{equation} 
\left( {{\partial _t} + {\bf{v}} \cdot \nabla } \right) \psi  =  - \eta {j_z} + {E_0} \, , \label{e1}
\end{equation}
and the vorticity equation
\begin{equation} 
\left( {{\partial _t} + {\bf{v}} \cdot \nabla } \right) {\omega_z} =  {\bf{B}} \cdot \nabla {j_z} + \nu {\nabla ^2}{\omega _z} \, . \label{e2}
\end{equation}
Here, $j_z = - \nabla^2 \psi$ is the electric current density and $\omega_z = \nabla^2 \phi$ is the plasma vorticity, both in the $z$ direction. The kinematic viscosity can be recognized as $\nu$, whereas $\eta$ stands for the electrical resistivity. The term $E_0 = \eta j_{z}^{(0)}$ represents an externally applied electric field, which is required to maintain the equilibrium magnetic field in the presence of a small but finite electrical resistivity.

This system of equations is dimensionless, with the lengths normalized to a convenient scale length $L$, the magnetic field to a convenient scale field strength $B_0$, and the time to the Alfv\'en time $\tau_A = L/v_A$, where $v_A$ is the Alfv\'en speed based on $B_0$. 
Therefore, the normalized resistivity corresponds to the inverse of the Lundquist number $\eta  = {D_\eta}/({v_A}L)$, being ${D_\eta}$ the magnetic diffusivity, while the normalized kinematic viscosity is linked to the magnetic Prandtl number by means of the relation $\nu = P_m \eta$.

We consider a slab plasma which is periodic in the $y$ direction and is bounded by perfectly conducting walls at $x=\pm 1$ (note that the normalizing length $L$ is chosen to be the half distance between the walls). Moreover, we consider a tearing-stable equilibrium given by
\begin{equation}
{\psi ^{(0)}}(x) =  - {x^2}/2  \, , \qquad {\phi ^{(0)}}(x) = 0 \, . \label{equil}
\end{equation}
This gives also that ${B_y}^{(0)}(x) = x$, ${j_z}^{(0)}(x) = 1$, and ${B_x}^{(0)}(x) = {v_x}^{(0)}(x) = {v_y}^{(0)}(x) = {\omega_z}^{(0)}(x) = 0$ (the normalizing field strength $B_0$ is chosen to be the in-plane equilibrium magnetic field at $\left| x \right| = 1$).

Now we suppose that the conducting walls are subject to a sudden displacement 
\begin{equation}
x_w =  \pm 1 \mp {\Xi (t)}\cos (ky) \ ,
\end{equation}
where $k=2\pi/ L_y$ and $\Xi (t) = {\Xi _0}\left( {1 - {e^{ - t/\tau }} - (t/\tau ){e^{ - t/\tau }}} \right)$ for $t \geq 0$ \cite{Fitz_2004a,Fitz_2004b}. The perturbation time scale $\tau$ is assumed to be significantly larger than the Alfv\'en time $\tau_A$ (thus the plasma can be considered in magnetostatic equilibrium everywhere except near the resonant surface at $x=0$) but much shorter than any characteristic reconnection time scale. Furthermore, the displacement is assumed to be small compared to the half distance between the walls, i.e. ${\Xi_0} \ll 1$. 
Consequently, the boundary conditions at the flux conserving walls can be written as $\psi ( \pm 1,y,t) =  - 1/2 - \Psi (t)\cos (ky)$, $ j_z( \pm 1,y,t) = 1$, $\phi ( \pm 1,y,t) =  \pm d/dt(\Xi (t)) {\sin (ky)}/k$ and $\omega_z( \pm 1,y,t) = 0$.
Note that $\Psi (t) = \Xi (t)$ since $d{\psi ^{(0)}}/{\left. {dx} \right|_{x= -1}} = 1$. For this reason we have also that ${\Psi_0} = {\Xi_0}$.

In Taylor's problem, as well as in tearing mode theory, it is convenient to divide the spatial domain into two regions: an ``outer region'', where non-ideal effects and plasma inertia can be neglected, and a thin ``inner region'' centered on the resonant surface at $x=0$, where resistivity, viscosity, and plasma inertia can be important. Hence, in the outer region the vorticity equation reduces to
\begin{equation}
{\bf{B}} \cdot \nabla {j_z} = 0 \, . \label{Ideal_Equil}
\end{equation}
Since the boundary perturbation is relatively small, we can write ${\psi_{{\text{out}}}} = {\psi ^{(0)}} + \psi_{{\text{out}}}^{(1)}$, where $\psi_{{\text{out}}}^{(1)}(x,y,t)$ is the perturbation in the form $\psi_{{\text{out}}}^{(1)}(x,y,t) = \psi _{{\text{out}}}^{(1)}(x,t)\cos (ky)$. Therefore, the linear approximation of Eq. (\ref{Ideal_Equil}) yields
\begin{equation}
x\left( {\partial_x^2\psi _{{\text{out}}}^{(1)} - {k^2}\psi_{{\text{out}}}^{(1)}} \right) = 0 \, . \label{Tearing_Eq}
\end{equation}
The solution of Eq. (\ref{Tearing_Eq}) which has parity $\psi_{{\text{out}}}^{(1)}(x,t) = \psi_{{\text{out}}}^{(1)}( - x,t)$ and satisfies the boundary condition $\psi_{{\text{out}}}^{(1)}(\pm 1,t) = {\left. {d{\psi ^{(0)}}/dx} \right|_{x = -1}}\Xi (t)$ is \cite{HK_1985}
\begin{equation}
\psi_{{\text{out}}}^{(1)}(x,t) = \psi_{{\text{out}}}^{(1)}(0,t)\left( {\cosh kx - \frac{{\sinh \left| {kx} \right|}}{{\tanh k}}} \right) + {\Psi}(t)\frac{{\sinh \left| {kx} \right|}}{{\sinh k}}  \, . \label{HK_eq_solution}
\end{equation}
In the vicinity of the resonant surface, namely around $x = 0$, the previous expression can be described by a Taylor expansion as
\begin{equation} \label{expansion_psi}
\psi _{{\text{out}}}^{(1)}(x,t) = \psi _{{\text{out}}}^{(1)}(0,t) + \frac{1}{2}\Delta {\psi _{{\text{out}}}}(t)\left| x \right| + \mathcal{O}\left( {{x^2}} \right) \, , 
\end{equation}
with $\Delta {\psi_{{\text{out}}}}$ indicating the gradient discontinuity
\begin{equation}  \label{delta_psi_def}
\Delta {\psi _{{\text{out}}}}(t) = {\left. {{\partial _x}\psi _{{\text{out}}}^{(1)}} \right|_{{0^ + }}} - {\left. {{\partial _x}\psi _{{\text{out}}}^{(1)}} \right|_{{0^ - }}} =  \Delta '_0  \psi _{{\text{out}}}^{(1)}(0,t) + \Delta '_s \Psi (t) \, ,
\end{equation}
where
\begin{equation} \label{Ess_def}
\Delta '_0 = - \frac{{2k}}{{\tanh k}} 
\end{equation}
is the standard tearing stability parameter \cite{FKR_1963}, which is negative in Taylor's problem, while
\begin{equation} \label{Ess_Esw_def}
\Delta '_s = \frac{{2k}}{{\sinh k}}
\end{equation}
parametrizes the contribution of the external source perturbation to the gradient discontinuity.

In order to determine the explicit evolution of the system, we have to solve an initial value problem and to match the solutions obtained in the outer and inner regions. The early evolution of the reconnected magnetic flux can be calculated in the linear approximation, as was done by Hahm and Kulsrud for the case in which $\nu / \eta \ll 1$ \cite{HK_1985} and by Fitzpatrick for the opposite case with $\nu / \eta \gg 1$ \cite{Fitz_2003}. In the next section we revisit their linear results, whereas in the subsequent sections we investigate the nonlinear evolution by performing both analytical and numerical calculations.

\section{Linear Theory}  \label{sec3}

The time dependence of the reconnected flux and the associated reconnection rate can be obtained by asymptotically matching the inner and outer solutions in the overlapping domain. With this aim in mind, we now consider the vorticity equation and Ohm's law in the inner region centered on the resonant surface $x=0$. Linearizing Eqs. (\ref{e1}) and (\ref{e2}) by considering perturbations in the form ${\psi ^{(1)}}(x,y,t) = {\psi ^{(1)}}(x,t)\cos (ky)$ and ${\phi ^{(1)}}(x,y,t) = {\phi ^{(1)}}(x,t)\sin (ky)$, and assuming that in the inner region ${k^2} \ll \partial_x^2$, we get
\begin{equation} 
{\partial _t}{\psi ^{(1)}} + kx{\phi ^{(1)}} = \eta \partial _x^2{\psi ^{(1)}}  \, , \label{e1_lin}
\end{equation}
\begin{equation} 
{\partial _t}\partial _x^2{\phi ^{(1)}} = kx\partial _x^2{\psi ^{(1)}} + \nu \partial _x^4{\phi ^{(1)}}  \, . \label{e2_lin}
\end{equation}
Performing the change of variable $\hat x \equiv kx$ and then Laplace trasforming and Fourier transforming the perturbed quantities as
\begin{equation} 
{\overline \psi  (\hat x,g)} = \int_0^\infty  {{\psi ^{(1)}}(\hat x,t)} \,{e^{ - gt}}dt   \,
\end{equation}
and
\begin{equation} 
\tilde \psi (\theta ,g) = \int_{ - \infty }^\infty  {\overline \psi  (\hat x,g)} \,{e^{ - i\theta \hat x}} d \hat x   \, ,
\end{equation}
the set of equations (\ref{e1_lin})-(\ref{e2_lin}) can be rewritten as
\begin{equation} 
g\tilde \psi  + i{\partial _\theta }\tilde \phi  =  - \eta {k^2}{\theta ^2}\tilde \psi  \, , \label{e1_lin_B}
\end{equation}
\begin{equation} 
g{k^2}{\theta ^2}\tilde \phi  = i{k^2}{\partial _\theta }({\theta ^2}\tilde \psi ) - \nu {k^4}{\theta ^4}\tilde \phi  \, . \label{e2_lin_B}
\end{equation}
Then, eliminating $\tilde \psi $ from Eq. (\ref{e2_lin_B}) by means of Eq. (\ref{e1_lin_B}), we obtain the layer equation
\begin{equation}  \label{layer_eq}
{\partial_\theta }\left( {\frac{{{\theta ^2}}}{{g + \eta {k^2}{\theta ^2}}}{\partial _\theta }\tilde \phi } \right) - g{\theta ^2}\tilde \phi  - \nu {k^2}{\theta ^4}\tilde \phi  = 0  \, .
\end{equation}
This equation must be solved subject to the condition 
\begin{equation}  \label{BC_theta_1}
\mathop {\lim }\limits_{\theta \to \infty } \tilde \phi \left( {\theta ,g} \right) = 0   \, . 
\end{equation}
In addition, the inner solution has to match the outer solution at (relatively) large $x$, i.e. small $\theta$. This boundary condition can be obtained by neglecting resistivity in the linearized form of Eq. (\ref{e1}), and substituting ${\psi _{{\text{out}}}^{(1)}}(x,t)$ with its Taylor expansion around $x = 0$. Then, after using the Fourier-Laplace transform, the boundary condition for the inner solution becomes 
\begin{equation}  \label{BC_theta_2}
\tilde \phi (\theta ,g)  \to  i \, \frac{g}{2} \, \overline \psi_{{\text{out}}}(0,g) \left( {\frac{\Delta (g)}{{\pi k\theta }} + 1 + \mathcal{O}\left( \theta  \right)} \right) 
\end{equation}
in the small $\theta$ limit, where we have defined 
\begin{equation}  \label{}
\Delta (g) \equiv \frac{{{{\overline {\Delta \psi } }_{{\text{out}}}}(g)}}{{{{\overline \psi }_{{\text{out}}}}(0,g)}} \, .
\end{equation}

The quantity $\Delta (g)$ can be evaluated by solving Eq. (\ref{layer_eq}) subject to the boundary conditions (\ref{BC_theta_1}) and (\ref{BC_theta_2}). Then, $\Delta (g)$ may be used in the Laplace transform of Eq. (\ref{delta_psi_def}) to obtain the ``outer'' magnetic flux
\begin{equation}  \label{outer_magnetic_flux}
{{\overline \psi }_{{\text{out}}}}(0,g) = \frac{{{\Delta '_s}\overline \Psi  (g)}}{{\Delta (g) - {\Delta '_0}}} \, .
\end{equation}

The reconnected flux can be calculated by considering the Laplace trasform of Eq (\ref{e1_lin}), which yields
\begin{equation}  \label{}
g \overline \psi (\hat x,g) =  - \hat x \overline \phi (\hat x,g) + \eta {k^2}\partial _{\hat x}^2 \overline \psi (\hat x,g) \, .
\end{equation}
By means of the Fourier transform we can rewrite this equation as
\begin{equation}  \label{}
g\overline \psi (\hat x,g) =  - \hat x \overline \phi (\hat x,g) - \eta {k^2}\int_{ - \infty }^\infty  {{\theta ^2} \tilde  \psi (\theta ,g){e^{i\theta \hat x}}} d\theta   \, .
\end{equation}
Then, using Eq. (\ref{e1_lin_B}) to eliminate $\tilde \psi$ and evaluating this equation at $x=0$, we obtain the Laplace transformed reconnected flux
\begin{equation}  \label{Eq_Rec_Flux_Lapl}
\overline \psi (0,g) = \frac{{2\eta {k^2}}}{g}\int_0^\infty  {\frac{{{i \theta ^2}}}{{g + \eta {k^2}{\theta ^2}}}{\partial _\theta }\tilde \phi \, d\theta }   \, .
\end{equation}

Finally, the inverse Laplace transform of $\overline \psi (0,g)$ gives us the time evolution of the reconnected flux ${\psi^{(1)}}(0,t)$, whose time derivative is a measure of the reconnection rate.

In the following we solve the layer Eq. (\ref{layer_eq}) and calculate the reconnection rate in two different linear regimes of the reconnection process, the inertial regime and the visco-resistive one. Note that here we consider a visco-resistive plasma which is characterized by plasma resistivity $\eta$ that is smaller than kinematic viscosity $\nu$. The opposite case can be deduced following the same procedure.

\subsection{Inertial phase}  \label{sec3.1}

We first consider the inertial phase, which occurs for \cite{Fitz_2003} $t \ll \tau_i$, with  $\tau_i \equiv {\nu ^{-1/3}}{k^{-2/3}}$. In this regime, resistivity and viscosity can be neglected in the layer equation, which reduces to
\begin{equation}  \label{layer_eq_inertial}
\partial _\theta ^2\tilde \phi  + \frac{2}{\theta }{\partial _\theta }\tilde \phi  - {g^2}\tilde \phi  = 0  \, .
\end{equation}
The solution of this equation that satisfies the boundary conditions (\ref{BC_theta_1}) and (\ref{BC_theta_2}) is
\begin{equation}  \label{SOL_layer_eq_inertial}
\tilde \phi (\theta,g) = i \, {{\overline \psi}_{{\text{out}}}}(0,g) \frac{{e^{-g\theta}}}{2\theta}  \, .
\end{equation}
Substituting this solution into Eq. (\ref{BC_theta_2}) and taking the small $\theta$ limit gives us
\begin{equation}  \label{Delta_inertial}
\Delta (g) =  - \frac{{k\pi }}{g} \, .
\end{equation}
Furthermore, substituting the solution (\ref{SOL_layer_eq_inertial}) into Eq. (\ref{Eq_Rec_Flux_Lapl}) and recalling that in this regime $\eta k^2 \theta^2 \ll g$, we get
\begin{equation}  \label{}
\overline \psi (0,g) = - \frac{{2\eta {k^2}}}{{{g^3}}} {\overline \psi_{{\text{out}}}}(0,g)  \, .
\end{equation}
Note that since $\left| { - 2\eta {k^2}/{g^3}} \right| \ll 1$, the inertial phase is a non-constant-$\psi$ regime. We will see in the next sections that this may have important implications in the nonlinear evolution of the reconnection process.

We can now evaluate ${\overline \psi_{{\text{out}}}}(0,g)$ by means of Eq. (\ref{outer_magnetic_flux}). Neglecting the small initial transient $\tau$ and also $\Delta '_0$ with respect to $\Delta (g)$, we obtain
\begin{equation}  \label{}
{\overline \psi} (0,g) = \frac{2}{\pi} \frac{{\eta k}}{{{g^3}}}{\Delta '_s} \Psi_0 \, .
\end{equation}
Finally, the inverse Laplace transform of this expression gives us the time evolution of the reconnected flux
\begin{equation}  \label{}
\psi_i (t) \equiv {\psi^{(1)}}(0,t) = \eta k\frac{{{\Delta '_s}}}{\pi }{\Psi _0}{t^2}  \, ,
\end{equation}
while the reconnection rate in the inertial regime can be evaluated as 
\begin{equation}  \label{ReconnRate_IN}
{\partial_t}{\psi_i}(t) = \eta k \frac{{2 {\Delta '_s}}}{\pi }{\Psi _0}t  \, .
\end{equation}

Note that in the inertial phase the current density ${j_z^{(1)}}(0,t)$ (which is proportional to the reconnection rate) increases linearly in time, whereas the width of the current channel $\delta \sim g/k$ \cite{HK_1985} shrinks inversely with time since $g \sim t^{-1}$. This leads to the formation of a narrow current sheet at the resonant surface. The thinning of this sheet towards singularity (for $t \to \infty$) is prevented by resistive diffusion, which becomes important as the current density piles up.

\subsection{Visco-resistive phase}  \label{sec3.2}

We now consider the visco-resistive regime, which occurs when \cite{Fitz_2003} $t \gg \tau_{vr} \equiv {\nu^{1/3}}{(\eta k)^{-2/3}}$ and the magnetic island width $w$ is less than the linear layer width ${\delta_{\nu \eta}}$ (defined later in the paper). We can roughly follow Refs. \cite{Porcelli_1987,Brunetti_2014} to solve Eq. (\ref{layer_eq}) by asymptotic matching the solutions obtained for large $\theta$ and relatively small $\theta$. The large $\theta$ domain represents an inner sublayer in real space, whereas the relatively small $\theta$ region corresponds to an outermost sublayer in real space.

We first consider large values of $\theta$ such that $\theta > {(g/\eta k^2)^{1/2}}$. In this asymptotic interval, recalling that we are assuming $\nu > \eta$, Eq. (\ref{layer_eq}) may be approximated and cast into the form
\begin{equation}  \label{layer_eq_VR_1}
\partial _\theta ^2{\tilde \phi _ > } - {\varpi ^{ - 6}}{\theta ^4}{\tilde \phi _ > } = 0  \, ,
\end{equation}
where ${{\varpi}^{-6}} \equiv \eta \nu {k^4}$ and $\tilde \phi_> (\theta, g)$ indicates the function that approximate $\tilde \phi (\theta, g)$ for $\theta > {(g/\eta k^2)^{1/2}}$. The analytical solution of this equation can be easily found by transforming it into a familiar form. To this aim,  
let us introduce the change of variable $\tilde \phi_> \equiv {\theta ^{1/2}} \tilde \chi $, which leads to
\begin{equation}  \label{layer_eq_VR_2}
{\theta ^2}\frac{{{d^2}\tilde \chi }}{{d{\theta ^2}}} + \theta \frac{{d\tilde \chi }}{{d\theta }} - \left( {\frac{1}{4} + {\varpi ^{ - 6}}{\theta ^6}} \right)\tilde \chi  = 0  \, .
\end{equation}
Then, with one more change of variable, precisely $\hat z \equiv {\theta ^3}/\left( {3{\varpi^3}} \right)$, we can write the previous equation in the form of the modified Bessel equation
\begin{equation}  \label{layer_eq_VR_3}
{\hat z^2}\frac{{{d^2}\tilde \chi }}{{d{{\hat z}^2}}} + \hat z\frac{{d\tilde \chi }}{{d\hat z}} - \left( {{{\hat z}^2} + {\alpha ^2}} \right)\tilde \chi  = 0  \, ,
\end{equation}
with $\alpha=1/6$. The general solution of this equation is $\tilde \chi = {c_1}{I_\alpha }(\hat z) + {c_2}{K_\alpha }(\hat z)$, where $I_{\alpha}$ and $K_{\alpha}$ are the $\alpha$-order modified Bessel functions of the first and second kind, respectively, while $c_1$ and $c_2$ can be determined from boundary conditions. Since $I_{\alpha}(\hat z)$ grows exponentially while $K_{\alpha}(\hat z)$ decays exponentially, the solution of Eq. (\ref{layer_eq_VR_1}) that vanishes for $\theta \to  + \infty$ is
\begin{equation}  \label{Phi_inner1a}
\tilde \phi_>  = {c_2} {\theta^{1/2}}{K_{1/6}}\left( {\frac{{{\theta^3}}}{{3{{\varpi}^3}}}} \right) \, .
\end{equation}

In the opposite case of relatively small values of $\theta$ such that $\theta < {\left( {\nu \eta {k^4}} \right)^{-1/6}}$, Eq. (\ref{layer_eq}) may be approximated as
\begin{equation}  \label{layer_eq_VR_5}
{\partial _\theta }\left( {\frac{{{\theta ^2}}}{{g + \eta {k^2}{\theta ^2}}}{\partial _\theta }\tilde \phi_< } \right) = 0  \, ,
\end{equation}
where $\tilde \phi_< (\theta, g)$ indicates the function that approximate $\tilde \phi (\theta, g)$ in this asymptotic interval. The general solution of this equation is ${{\tilde \phi }_ < } = ( - g/\theta  + \eta k^2 \theta )c_3 + c_4$, therefore, by applying the boundary condition (\ref{BC_theta_2}) we obtain
\begin{equation}  \label{Phi_inner2a}
{{\tilde \phi }_ < } = \left( { - \frac{g}{\theta } + \eta {k^2}\theta } \right) i \frac{{\Delta (g)}}{{2\pi k}}{{\overline \psi }_{{\text{out}}}}(0,g) - i \frac{{g}}{2}{{\overline \psi }_{{\text{out}}}}(0,g)  \, .
\end{equation}

The quantity $\Delta (g)$ may be evaluated by matching the functions $\tilde \phi_> $ and $\tilde \phi_< $ in the overlapping interval ${(g/\eta )^{1/2}}/k <  \theta <  {\left( {\nu \eta {k^4}} \right)^{ - 1/6}}$. In this region $\tilde \phi_< $ becomes
\begin{equation}  \label{Phi_inner_2matching}
{{\tilde \phi }_ < } = i\frac{g}{2}\,{{\overline \psi }_{{\text{out}}}}(0,g)\left( {\frac{{\eta k\theta \Delta (g)}}{{g\pi }} - 1} \right) \, ,
\end{equation}
while the expansion of $\tilde \phi_>$ that match the previous solution is
\begin{equation}  \label{Phi_inner_1matching}
{\tilde \phi }_ > =  i \frac{g}{2}\,{{\overline \psi }_{\text{out}}}(0,g) \left( {{6^{2/3}}\frac{{\Gamma \left( {\frac{5}{6}} \right)}}{{\Gamma \left( {\frac{1}{6}} \right)}}\frac{\theta }{\varpi } -1} \right)  \, ,
\end{equation}
where $\Gamma$ indicates the Gamma function. Therefore, the matching between $\tilde \phi_> $ and $\tilde \phi_< $ gives us
\begin{equation}  \label{Delta_vr}
\Delta (g) = g \, \tau_* \, ,  
\end{equation}
where we have defined
\begin{equation}
\tau_*  \equiv  \pi c_*  \frac{{{\nu^{1/6}}}}{{{k^{1/3}}{\eta^{5/6}}}}  \quad {\text{and}}  \quad  c_*  \equiv  {6^{2/3}}\frac{{\Gamma \left( {\frac{5}{6}} \right)}}{{\Gamma \left( {\frac{1}{6}} \right)}}  \, .
\end{equation}

$\Delta (g)$ allows us to evaluate the reconnected flux, which is equal to the outer flux in the visco-resistive regime. In fact, recalling that in this regime $g \ll \eta k^2 \theta^2$, Eq. (\ref{Eq_Rec_Flux_Lapl}) gives us
\begin{equation}  \label{}
\overline \psi (0,g) = {\overline \psi _{{\text{out}}}}(0,g) \, .
\end{equation}
Then, neglecting the small initial transient $\tau$, from Eq. (\ref{outer_magnetic_flux}) we get
\begin{equation}  \label{}
{\overline \psi}(0,g) = \frac{{{\Delta '_s}{\Psi _0}}}{{{\tau _*}\,{g^2} - {\Delta '_0}g}}  \, ,
\end{equation}
which can be inverse Laplace transformed to obtain
\begin{equation}  \label{ReconnFlux_VR}
\psi_{vr}(t) \equiv {\psi^{(1)}}(0,t) = \frac{{{\Delta '_s}{\Psi_0}}}{{{\Delta '_0}}}\left( {{e^{{\Delta '_0}t/{\tau_*}}} - 1} \right)  \, .
\end{equation}
Therefore, the reconnection rate in the visco-resistive regime is
\begin{equation}  \label{ReconnRate_VR}
{\partial_t}{\psi_{vr}}(t) =  \Delta '_s \frac{{{\Psi_0}}}{{{\tau_*}}} e^{{\Delta '_0}t/{\tau_*}}  \, .
\end{equation}

\section{Nonlinear Theory} \label{sec4}

The linear tearing mode analysis breaks down and nonlinearities must be considered when the magnetic island width grows to a size comparable or exceeding the linear layer width, which for a visco-resistive plasma corresponds to \cite{Porcelli_1987,Fitz_NF1993,Fitz_2003,Grasso_2008}
\begin{equation}  \label{}
{\delta_{\nu \eta}} = 4{\left( {\pi {c_*}} \right)^{1/2}}\frac{{{{\left( {\nu \eta } \right)}^{1/6}}}}{{{k^{1/3}}}} \, .
\end{equation}

In particular, we will see that three possible nonlinear scenarios may occur.

i) A Rutherford evolution \cite{Rutherford_1973} takes place if the magnetic island that enters into the nonlinear regime is characterized by a constant-$\psi$ behavior.

ii) A Sweet-Parker-like evolution \cite{Waelbroeck_1989} happens if the magnetic island that passes into the nonlinear regime has a non-constant-$\psi$ behavior and the current sheet is stable. This phase then gives way (on the time scale of resistive diffusion) to the Rutherford regime.

iii) A ``plasmoid phase'' occurs if the magnetic island that enters into the nonlinear regime has a non-constant-$\psi$ behavior and the current sheet is unstable.

In the following we analyze all these nonlinear scenarios and show how they depend on both the microscopic ($\nu$ and $\eta$) and macroscopic ($\Psi$, $\Delta '_s$, $\Delta '_0$ and $k$) parameters of the Taylor problem. While the first two scenarios where discussed for the first time in the pioneer works by Hahm and Kulsrud \cite{HK_1985} and Wang and Bhattacharjee \cite{WB_1992} (and reconsidered later by Fitzpatrick \cite{Fitz_2003} in the large magnetic Prandtl number limit), the plasmoid-dominated phase of the Taylor problem is entirely new. Since only part of the analysis of the visco-resistive plasmoid phase is strictly related to the Taylor paradigm, the applicability of the results is much more general.

\subsection{Rutherford phase}  \label{sec4.1}

As a result of the fact that the forced reconnection process evolves initially according to the inertial phase (which is a non-constant-$\psi$ regime), the system proceeds in the nonlinear regime in agreement to a Rutherford evolution only if the inertial phase gives way to a visco-resistive phase (which is a constant-$\psi$ regime) before the transition to the nonlinear regime. Therefore, if we suppose that the transition from the non-constant-$\psi$ to the constant-$\psi$ magnetic island occurs at $t \sim \tau_{vr}$, the existence of the visco-resistive phase requires that \cite{Fitz_2003}
\begin{equation}  \label{Rutherford_Condition}
w(\tau_{vr}) \ll {\delta_{\nu \eta}} \, .
\end{equation}
Recalling that the width of constant-$\psi$ islands is linked to the reconnected flux by means of the relation
\begin{equation}  \label{COST_PSI_ISLAND}
w(t) = 4 \sqrt{{\psi ^{(1)}}(0,t)}  \, ,
\end{equation}
we can specify condition (\ref{Rutherford_Condition}) by evaluating the reconnected flux at $\tau_{vr}$. Note that in Eq. (\ref{COST_PSI_ISLAND}) the contribution of the equilibrium current density does not appear explicitly since in this problem, without loss of generality, we have $d^2{\psi^{(0)}}/{\left. {dx^2} \right|_{x=0}} = 1$. An estimation of the reconnected flux at $\tau_{vr}$ can be obtained from Eq. (\ref{ReconnFlux_VR}), which gives us
\begin{equation}  \label{}
{\psi^{(1)}}(0,\tau_{vr}) \sim  \frac{{{\Delta '_s}{\Psi _0}}}{{{\Delta '_0}}}\left( {{e^{{\Delta '_0} \tau_{vr}/{\tau_*}}} - 1} \right)  \, .
\end{equation}
Since $\tau _{vr} \ll \tau_*$, the previous relation may be further semplified by using the Taylor series expansion of the exponential function. Hence, we can estimate
\begin{equation}  \label{}
w(\tau_{vr}) \sim 4 \sqrt{\Delta '_s \Psi_0 \frac{\tau_{vr}}{\tau_*}}   \, .
\end{equation}
Finally, condition (\ref{Rutherford_Condition}) can be rewritten as
\begin{equation}  \label{Rutherford_Condition_2}
{\Psi _0} \ll \frac{{{{\left( {\nu \eta } \right)}^{1/6}}}}{{{\Delta '_s}{k^{1/3}}}} \, ,
\end{equation}
which tells us that the reconnection process proceeds in the nonlinear regime according to a Rutherford evolution only if the magnetic perturbation amplitude is sufficiently small (i.e., for a sufficiently small displacement $\Xi_0 = \Psi_0 / {\left( {d{\psi ^{(0)}}/{{\left. {dx} \right|}_{x= -1}}} \right)}$).

We now proceed to determine the reconnection rate in the Rutherford phase, i.e., assuming that condition (\ref{Rutherford_Condition_2}) is satisfied. In this case, since the constant-$\psi$ approximation implies that $\psi (x,y,t) = {\psi ^{(0)}}(x) + {\psi ^{(1)}}(0,t)\cos (ky)$, the magnetic island geometry is entirely determined in terms of its width, whose evolution follows the Rutherford equation
\begin{equation}  \label{Rutherford_Eq}
\frac{{\cal I}}{\eta }\frac{{dw}}{{dt}} = \Delta '_0 + \Delta '_s  {\Psi}(t) {\left( {\frac{4}{w}} \right)^2} \, ,
\end{equation}
where ${\cal I}=0.823$ \cite{Rutherford_1973,Fitz_NF1993}. Here we may again neglect the perturbation time scale $\tau$ since it is assumed to be much shorter than any characteristic reconnection time scale. Performing the change of variables $\hat t \equiv t/ \tau_\wedge$ and $\hat w \equiv w/ w_\wedge$, with
\begin{equation}  \label{}
\tau_\wedge = \frac{{4 {\cal I}}}{{\eta ( -\Delta '_0)}}\sqrt {\frac{{{\Delta '_s}{\Psi _0}}}{{(-\Delta '_0)}}}  \, ,
\end{equation}
\begin{equation}  \label{}
w_\wedge = 4 \sqrt {\frac{\Delta '_s \Psi_0}{( - \Delta '_0)}} \, ,
\end{equation}
we can rewrite Eq. (\ref{Rutherford_Eq}) as
\begin{equation}  \label{Eq_hatw}
\frac{d \hat w}{d \hat t} = \frac{1}{{\hat w}^2} - 1  \, .
\end{equation}
From this equation we can see that $\hat w$ increases in time and reaches $\hat w \sim 1$ when $t \sim \tau_\wedge$. Its integration gives that \cite{HK_1985}
\begin{equation}  \label{Sol_Eq_hatw}
\hat w =  \tanh (\hat w + \hat t)   \, .
\end{equation}
Then, the time evolution of the reconnected flux can be easily calculated in two interesting limits of Eq. (\ref{Sol_Eq_hatw}). For $t \ll \tau_\wedge$ we may expand $\tanh (\hat w + \hat t)$ into its Taylor series, which leads us to obtain
\begin{equation}  \label{}
\psi_R (t) = \frac{{{\Delta '_s}{\Psi _0}}}{{( - \Delta '_0)}}{\left( {\frac{3t}{{\tau_\wedge}}} \right)^{2/3}}  \quad {\text{for}} \quad  t \ll {\tau_\wedge}  \, .
\end{equation}
In the opposite case $t \gg \tau_\wedge$, we may neglect $\hat w$ with respect to $\hat t$ in the argument of the hyperbolic tangent, so that the time evolution of the reconnected flux becomes
\begin{equation}  \label{}
\psi_R (t) = \frac{{{\Delta '_s}{\Psi _0}}}{{( - \Delta '_0)}}{\tanh ^2}\left( {\frac{t}{{{\tau_\wedge}}}} \right) \quad {\text{for}} \quad  t \gg \tau_\wedge \, .
\end{equation}

It is now straightforward to evaluate the reconnection rate as 
\begin{equation}  \label{ReconnRate_RU1}
{\partial _t}{\psi _R}(t) = \frac{{2{\Delta '_s}{\Psi _0}}}{{( - \Delta '_0){\tau_\wedge}}}{\left( {\frac{{3t}}{{{\tau_\wedge}}}} \right)^{-1/3}} \quad {\text{for}} \;  t \ll \tau_\wedge \, ,
\end{equation}
\begin{equation}  \label{ReconnRate_RU2}
{\partial _t}{\psi _R}(t) = \frac{{2{\Delta '_s}{\Psi _0}}}{{( - \Delta '_0){\tau_\wedge}}}\tanh \left( {\frac{t}{{{\tau_\wedge}}}} \right){\operatorname{sech} ^2}\left( {\frac{t}{{{\tau_\wedge}}}} \right)  \quad {\text{for}} \;  t \gg \tau_\wedge \, .
\end{equation}
Note that in the Rutherford regime the magnetic reconnection process is slow and depends strongly on the plasma resistivity. However, in the following we show that the evolution of the reconnection process may be very different for larger perturbation amplitudes.

\subsection{Sweet-Parker-like phase}  \label{sec4.2}

An elongated current sheet characterizes the initial stage of the nonlinear reconnection process if the transition to the nonlinear regime occurs while the magnetic island has a non-constant-$\psi$ behaviour. In fact, non-constant-$\psi$ islands grow faster than the current can diffuse out of the reconnecting region, resulting in a strong current sheet at the resonant surface \cite{Waelbroeck_1989}. This nonlinear scenario occurs if \cite{Fitz_2003}
\begin{equation}  \label{SP_Condition}
w(\tau_{vr}) \gtrsim {\delta_{\nu \eta}} \, ,
\end{equation}
namely if a constant-$\psi$ magnetic island cannot form prior to the transition into the nonlinear regime. Following the arguments adopted to obtain condition (\ref{Rutherford_Condition_2}), inequality (\ref{SP_Condition}) can be rewritten as
\begin{equation}  \label{SP_Condition_2}
\Psi_0  \gtrsim \frac{{{{\left( {\nu \eta } \right)}^{1/6}}}}{{{\Delta '_s}{k^{1/3}}}} \equiv \Psi_{W} \, ,
\end{equation}
which shows that the nonlinear evolution of the reconnection process is characterized by an elongated current sheet as predicted in Ref. \cite{Waelbroeck_1989} if the perturbation amplitude is sufficiently large.

We now analyze the evolution of the reconnection process when condition (\ref{SP_Condition_2}) is satisfied and also assuming a stable current sheet. An unstable current sheet is instead considered in the next subsection, together with the conditions for the instability. Here we calculate the reconnection rate by performing a Sweet-Parker-like analysis in analogy with what was done by Wang and Bhattacharjee \cite{WB_1992}, but taking into account also plasma viscosity effects, which cannot be neglected if $\nu > \eta$.

As is customary in this type of analysis, we consider a reconnecting current sheet (diffusion region) with length $2 L_{cs}$ and width $2 \delta \ll 2 L_{cs}$. Then, assuming $\partial_y \ll \partial_x$ in the sheet and recalling that ${\bf{v}} \cdot \nabla \psi = 0$ at the $X$-point, the reconnection rate may be evaluated from Eq. (\ref{e1}) as
\begin{equation}  \label{Rec_rate_at_X_1}
{\left. {{\partial _t}{\psi ^{(1)}}} \right|_X} = \eta \left( {{{\left. {\partial _x^2{\psi ^{(1)}}} \right|}_X}{{\left. { + {\partial _y^2}{\psi ^{(1)}}} \right|}_X}} \right) \approx \eta \frac{B_{y,u}}{\delta}  \, ,
\end{equation}
where $B_{y,u}$ indicates the magnetic field upstream of the current sheet. Since in this problem $\partial_z = 0$, Faraday's law $\nabla  \times {\mathbf{E}} =  - {\partial _t}{\mathbf{B}}$ tells us that for a quasistationary magnetic field the $z$ component of the electric field is uniform. Therefore, ${\partial_t}{\psi^{(1)}} |_X$ can also be evaluated from Ohm's law in the ideal region just upstream of the reconnection layer
\begin{equation}  \label{Rec_rate_at_X_2}
{\left. {{\partial _t}{\psi ^{(1)}}} \right|_X} \approx {v_u}{B_{y,u}}  \, .
\end{equation}
Of course, $v_u$ indicates the plasma velocity upstream of the current sheet. Thus, from Eqs. (\ref{Rec_rate_at_X_1}) and (\ref{Rec_rate_at_X_2}) we can evaluate the width of the sheet as
\begin{equation}  \label{delta_SP}
\delta  \approx  {\left( {\eta \frac{L_{cs}}{v_d}} \right)^{1/2}} \, ,
\end{equation}
where we have estimated $v_u \approx v_d \delta /{L_{cs}}$ from mass conservation for an incompressible flow.

We have now to evaluate the downstream velocity $v_d$, which may be done from energy balance considerations \cite{Park_1984}. Assuming a linear increase of the outflow along the current sheet and a Couette flow between $x= \delta$ and the resonant surface $x=0$, we obtain
\begin{equation}  \label{en_balance_1}
\frac{1}{2} {B_{y,u}^2} \approx \frac{1}{2}v_d^2 + \nu \frac{{{v_d}/2}}{{\delta^2}}{L_{cs}} \, .
\end{equation}
Therefore, with the help of Eq. (\ref{delta_SP}), we can find
\begin{equation}  \label{en_balance_2}
v_d \approx B_{y,u} {\left( {1 + \frac{\nu }{\eta }} \right)^{-1/2}} \, .
\end{equation}
In the limit $\nu /\eta  \ll 1$ we recover the standard Sweet-Parker result, namely that the downstream velocity corresponds to the Alfv\'en speed based on the reconnecting component of the magnetic field upstream of the diffusion region, $v_d \approx B_{y,u}$. However, for increasing values of viscosity the downstream velocity decreases due to energy dissipation, and it approaches $v_d \approx B_{y,u} (\eta /\nu )^{1/2}$ for $\nu /\eta \gg 1$.

We can now substitute the expressions for $\delta$ and $v_d$ into Eq. (\ref{Rec_rate_at_X_1}) to get
\begin{equation}  \label{Rec_rate_at_X_3}
{\left. {\partial _t}{\psi ^{(1)}} \right|_X}  \approx  B_{y,u}^{3/2} L_{cs}^{-1/2} {\eta^{1/2}}{\left( {1 + \frac{\nu}{\eta}} \right)^{-1/4}}  \, ,
\end{equation}
as found by Park {\it et al.} \cite{Park_1984} in their extension of the Sweet-Parker model for visco-resistive MHD plasmas. However, here we want to go a step further by evaluating  $B_{y,u}$ and $L_{cs}$ (supposed to be known in Sweet-Parker type analysis) as a function of the parameters of the Taylor problem. To this aim, we use Eq. (\ref{delta_psi_def}) to express $B_{y,u}$ as a function of the external forcing. Neglecting the small initial transient and the saturation term involving $\Delta '_0$, we can write $B_{y,u} \approx \Delta '_s \Psi_0 /2$. Furthermore, numerical simulations indicate that the current sheet length is approximately one third of the periodicity length,
i.e. $2 L_{cs} \approx L_y/3 = 2\pi /3k$. Therefore, from Eq. (\ref{Rec_rate_at_X_3}) we obtain
\begin{equation}  \label{Rec_rate_at_X_4}
{\partial_t}{\psi_{SP}} \equiv {\left. {\partial _t}{\psi ^{(1)}} \right|_X} \approx \frac{1}{3} k^{1/2} {\left( {{\Delta '_s}{\Psi _0}} \right)^{3/2}} {\eta^{1/2}}{\left( {1 + \frac{\nu}{\eta}} \right)^{-1/4}} \, .
\end{equation}
Finally, integrating this relation we obtain the time evolution of the reconnected flux 
\begin{equation}  \label{RecFlux_SP}
\psi_{SP}(t) \equiv  {\left. {\psi }^{(1)} (t) \right|_X} \approx  \frac{1}{3} k^{1/2} {\left( {{\Delta '_s}{\Psi _0}} \right)^{3/2}} {\eta^{1/2}}{\left( {1 + \frac{\nu}{\eta}} \right)^{-1/4}} t \, .
\end{equation}

Note that even if the reconnection rate in this modified Sweet-Parker regime is higher than in the Rutherford one, it is still too slow to explain fast magnetic reconnection events in plasmas characterized by very small resistivity and viscosity. We will see in the following that in such cases the plasmoid instability may resolve this issue.

\subsection{Plasmoid phase}  \label{sec4.3}

The analysis of the previous section relies on the assumption that the reconnecting current sheet remains stable during the evolution of the system. However, it should be pointed out that this is a strong assumption, which may be violated for small values of plasma resistivity and viscosity. In fact, despite the stabilizing effect of the outflow \cite{Bulanov_1978}, we might expect a tearing instability to arise in very thin reconnecting current sheets. A large number of numerical simulations have indeed confirmed this prediction  \cite{Bisk_1986,Malara_1992,Lou_2005,Daugh_2006,Daugh_2007,Klimas_2008,BHYR_2009,SLUSC_2009,Daugh_2009,Cassak_2009,LCV_2010,Ni_2010,SC_2010,
Skender_2010,HB_2010,Barta_2011,Shen_2011,Loureiro_2012,Mei_2012,Baty_2012,Poye_2014,CGW_2014, 
Liu_2014,Militello_2014}. In particular, numerical simulations have shown that the tearing instability arises if the aspect ratio of the reconnecting current sheet exceeds a certain threshold, i.e. when
\begin{equation} \label{Plasmoid_Condition_1}
\delta  < \delta_c = \epsilon_c L_c  \, ,
\end{equation}
being $\epsilon_c$ the critical inverse aspect ratio, which was found to lie in the range \cite{Bisk_1986,BHYR_2009,SLUSC_2009,Cassak_2009,SC_2010,Skender_2010,HB_2010,Loureiro_2012,Baty_2012} ${\frac{1}{100}}$ - ${\frac{1}{200}}$. This relation tells us that the analysis presented in the previous subsection applies if the width of the reconnecting current sheet is larger than a critical width, namely if $\delta > \delta_c$. Here instead we focus on the opposite case $\delta < \delta_c$.

Let us first consider in more detail condition (\ref{Plasmoid_Condition_1}). Assuming a critical current sheet of width ${\delta_c} \approx {\left( {\eta {L_c}/{v_d}} \right)^{1/2}}$, we can estimate $L_c \approx \epsilon_c^{-2} \eta /{v_d}$. Then, we may use Eq. (\ref{en_balance_2}) to get
\begin{equation} \label{L_c}
{L_c} \approx \frac{\eta}{{\epsilon_c^{2}{B_{y,u}}}}{\left( {1 + \frac{\nu}{\eta}} \right)^{1/2}}  \, .
\end{equation}
In our normalized units $B_{y,u} = v_{A,u}$, where $v_{A,u}$ indicates the Alfv\'en speed upstream of the diffusion region. Therefore, we can rewrite the condition for the tearing instability of the (visco-resistive) reconnecting current sheet as
\begin{equation} \label{Plasmoid_Condition_2}
S = \frac{{{L_{cs}}{v_{A,u}}}}{\eta} > S_c = \frac{{{L_c}{v_{A,u}}}}{\eta} \approx  \epsilon_c^{-2} {\left( 1 + {\frac{\nu }{\eta }} \right)^{1/2}} \, ,
\end{equation}
where $S_c$ is the critical Lundquist number based on the critical length $L_c$. In the limit $\nu \ll \eta$ it follows that $S_c \approx \epsilon_c^{-2}$ \cite{BHYR_2009,Cassak_2009}, whereas in the limit $\nu \gg \eta$ we obtain $S_c \approx \epsilon_c^{-2} \left( {\nu/ \eta } \right)^{1/2}$, which agrees with the condition recently proposed by Loureiro {\it et al.} \cite{LSU_2013}.

\begin{figure}
\begin{center}
\includegraphics[bb = 0 4 359 233, width=8.6cm]{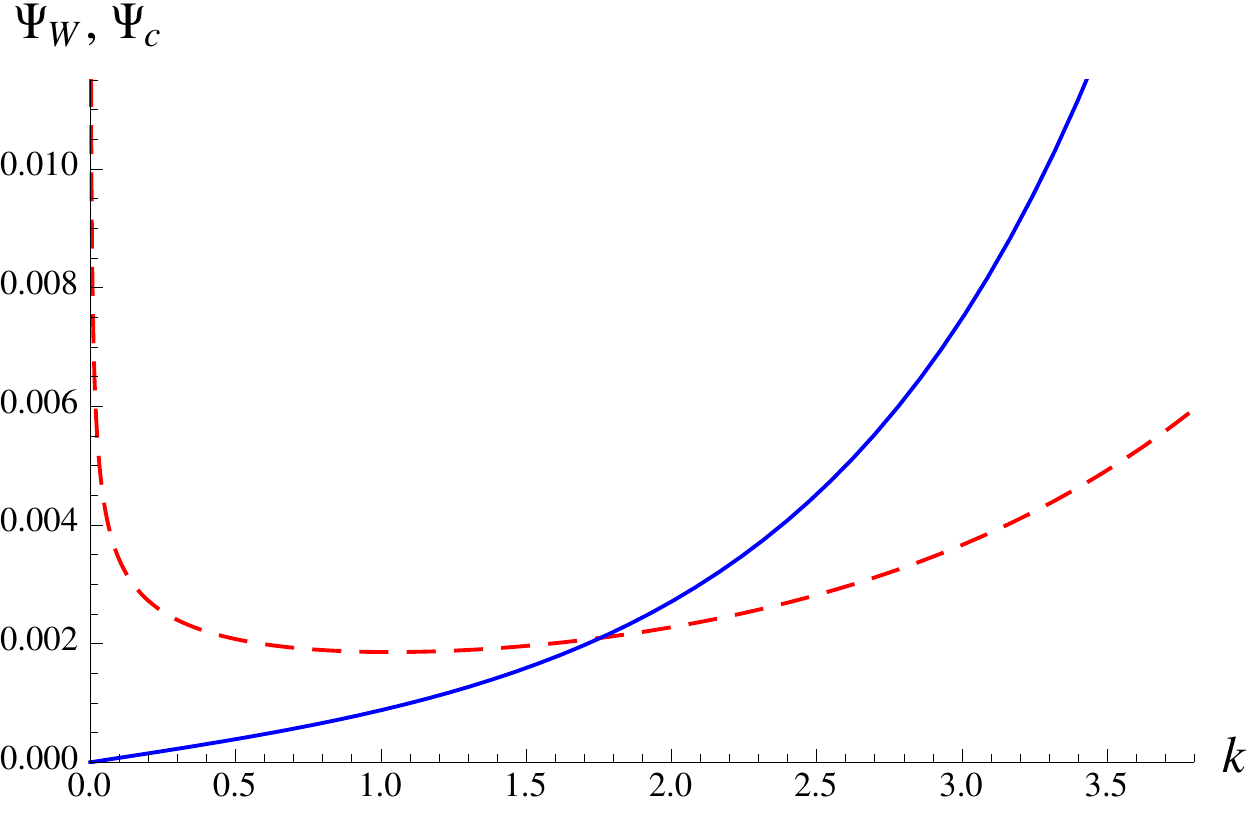}
\end{center}
\caption{${\Psi _{W}} = ({\Delta '_s}{k^{1/3}})^{-1} {\left( {\nu \eta } \right)^{1/6}}$ (red dashed line) and ${\Psi _c} = C \left( {k/{\Delta '_s}} \right) \eta {\left( 1 + {\nu/ \eta } \right)^{1/2}}$ (blue solid line) as a function of the perturbation wave number $k$ for $\eta = 10^{-8}$, $\nu = 10^{-7}$ and $C=2(150)^2$. The critical wave number $k^*$ is approximately given by the intersection of the two curves.}
\label{fig1}
\end{figure}

We are interested in a threshold condition for the perturbation amplitude rather than a condition on the Lundquist number based on the length of the current sheet. In order to achieve it, we recall that $B_{y,u} \approx {\Psi_0} {\Delta '_s}/2$ and estimate $L_{cs} \sim k^{-1}$. Therefore, we obtain the condition
\begin{equation}  \label{Plasmoid_Condition_3}
\Psi_0  >  C {\frac{k}{{{\Delta '_s}}}} \eta {\left( {1 + \frac{\nu }{\eta }} \right)^{1/2}} \equiv {\Psi_c} \, ,
\end{equation}
where $C \sim 2 \epsilon_c^{-2}$ is a  multiplicative constant. The fragmentation of the reconnecting current sheet occurs when both conditions (\ref{SP_Condition_2}) and (\ref{Plasmoid_Condition_3}) are satisfied. Hence, the plasmoid phase is triggered by the external perturbation when
\begin{equation}  \label{Plasmoid_Condition_4}
\Psi_0 
\begin{cases}
    > \Psi_c \, ,   &  \text{if } \Psi_c \gtrsim \Psi_{W} \, , \\
    \gtrsim \Psi_{W} \, ,    &  \text{if } \Psi_c < \Psi_{W} \, .
\end{cases}
\end{equation}
Since $\Psi_{W} \propto k^{-4/3} \sinh k$ and $\Psi_c \propto \sinh k$, there exists a critical wave number $k^*$ below which the reconnecting current sheet always leads to the plasmoid formation (see Fig. \ref{fig1}). Furthermore, since $k^*$ increases for decreasing values of $\eta$ and $\nu$, the disruption of the current sheet is unavoidable in plasmas with very small resistivity and viscosity. Note also that both $\Psi_{W}$ and $\Psi_c$ decrease with decreasing resistivity and viscosity, therefore the plasmoid phase may occur also for modest external perturbations if $\eta$ and $\nu$ are particularly small. 

We can now evaluate the reconnection rate in the plasmoid phase as the rate of change of the flux reconnected at the main $X$-point. In fact, as nicely observed in Ref. \cite{ULS_2010}, only the open-flux parcels matter when evaluating the total reconnection rate, since a perfectly closed plasmoid does not carry any net reconnected flux. Therefore, summing and subtracting the contributions of all the reconnection layers one can find that the total reconnection rate is given only by the reconnection region at the main $X$-point.

In a highly nonlinear regime the reconnection process is strongly time dependent, with plasmoids constantly being generated, ejected and merging with each other. However, we may assume a statistical steady state with a marginally stable current sheet located at the main $X$-point. 
In this case, the reconnection rate in statistical steady state can be evaluated as
\begin{equation}  \label{Rec_rate_plasmoids_at_X_1}
{\partial_t}{\psi_{p}} \equiv  \left\langle {{{\left. {{\partial _t}{\psi ^{(1)}}} \right|}_X}} \right\rangle  \approx \eta \frac{B_{y,u}}{\delta_c}  \, ,
\end{equation}
where $\left\langle  \ldots  \right\rangle $ denotes time average and 
\begin{equation}  \label{delta_c}
\delta_c \approx {\left( {\eta \frac{L_{c}}{v_d}} \right)^{1/2}}  \, .
\end{equation}
Evaluating the velocity downstream of the marginal current sheet by means of Eq. (\ref{en_balance_2}), we can estimate $\delta_c \approx \eta {\left( 1 + {\nu/ \eta } \right)^{1/2}} {\left( {{\epsilon_c}{B_{y,u}}} \right)^{-1}}$. Therefore, the resulting reconnection rate is
\begin{equation} \label{Rec_rate_plasmoids_at_X_2}
{\partial_t}{\psi_p} \approx \epsilon_c B_{y,u}^2 {\left( {1 + \frac{\nu }{\eta }} \right)^{-1/2}}  \, .
\end{equation}
This expression shows us that when plasmoid-dominated reconnection develops in a visco-resistive plasma, the reconnection rate decreases with increasing values of magnetic Prandtl number $P_m = \nu / \eta$. This behavior is very different from what can be found when $\nu \ll \eta$. In this latter case, indeed, the resulting reconnection rate in statistical steady-state becomes ${\partial_t}{\psi_p}  \approx  \epsilon_c B_{y,u}^2$ \cite{HB_2010}, and is therefore independent of the microscopic plasma parameters. However, as viscosity increases, this picture changes because the viscous energy dissipation leads to a decrease of the outflow velocity.

Finally, in order to evaluate the reconnection rate only as a function of the parameters of the Taylor problem, we estimate $B_{y,u}$ by means of Eq. (\ref{delta_psi_def}). Therefore, we obtain the relation
\begin{equation} \label{Rec_rate_plasmoids_at_X_3}
{\partial_t}{\psi_p} \approx \epsilon_c{\left( {{\Delta '_s}{\Psi_0}} \right)^2} {\left( {1 + \frac{\nu }{\eta }} \right)^{-1/2}} \, ,
\end{equation}
which shows that the reconnection rate in the plasmoid-dominated regime depends strongly on the external forcing. A strong dependence was found also for the stable current sheet case, as can be seen from Eq. (\ref{Rec_rate_at_X_4}). This is because the external forcing increases the magnetic field upstream of the reconnecting current sheet. A similar observation was made also for laser-driven magnetic reconnection by Fox {\it et al.} \cite{Fox_2011,Fox_2012}, who explained the fast reconnection rates observed in strongly laser-driven systems as the result of substantial upstream flux-pileup effects.

Note that the disruption of the current sheet increases significantly the reconnection rate with respect to the hypothetical case of an elongated current sheet that would remain stable even for very small values of resistivity and viscosity. Indeed, Eq. (\ref{Rec_rate_plasmoids_at_X_3}) shows us that the reconnection rate in the visco-resistive plasmoid phase depends on the magnetic Prandtl number $P_m = \nu / \eta$ but not on the Lundquist number $S$, leading to the possibility of attaining fast magnetic reconnection even at very large Lundquist numbers.

It is important to highlight that most of the presented theory applies to general reconnecting current sheets, irrespective of being generated by a strong external forcing (large $\Psi (t)$) or a spontaneous process in a strongly unstable equilibrium (large $\Delta '_0$), or a combination of both. Indeed, the condition (\ref{Plasmoid_Condition_1}) for the plasmoid instability of a reconnecting current sheet has general validity and extends the previously proposed conditions \cite{BHYR_2009,Cassak_2009,LSU_2013} to arbitrary ratios $\nu / \eta$. Similarly, also the relation (\ref{Rec_rate_plasmoids_at_X_2}) for the reconnection rate in the plasmoid-dominated regime is valid for both spontaneous and forced reconnection. In particular, this formula generalizes the previous results (${\partial_t}{\psi_p}  \approx  \epsilon_c B_{y,u}^2$ \cite{HB_2010} for $\nu \ll \eta$ and ${\partial_t}{\psi_p} \approx \epsilon_c B_{y,u}^2 (\eta/\nu)^{1/2}$ \cite{CGW_2014} for $\nu \gg \eta$) for arbitrary magnetic Prandtl number.

\section{Numerical Results} \label{sec5}

We show now the different scenarios of the Taylor problem by numerically solving the system of Eqs. (\ref{e1}) and (\ref{e2}) with equilibrium and boundary conditions as specified in Sec. \ref{sec2}. To this aim, we employ a numerical code that splits all the fields into the time-independent equilibrium and an evolving perturbation, which is advanced in time according to a third order Adams-Bashforth algorithm. A compact finite difference algorithm is adopted to compute the spatial operations in the $x$ direction (on a non-equispaced grid), while a pseudospectral method is used for the periodic direction. We adopt a space discretization of 512$\times$8192 grid points, so that the minimum step size in the $x$ direction is $d{x_{\min }} = 1.6 \times {10^{-3}}$, whereas in the $y$ direction we have $dy = 6.14 \times {10^{-3}}$.

\begin{figure}
\begin{center}
\includegraphics[bb =1 1 358 278, width=8.2cm]{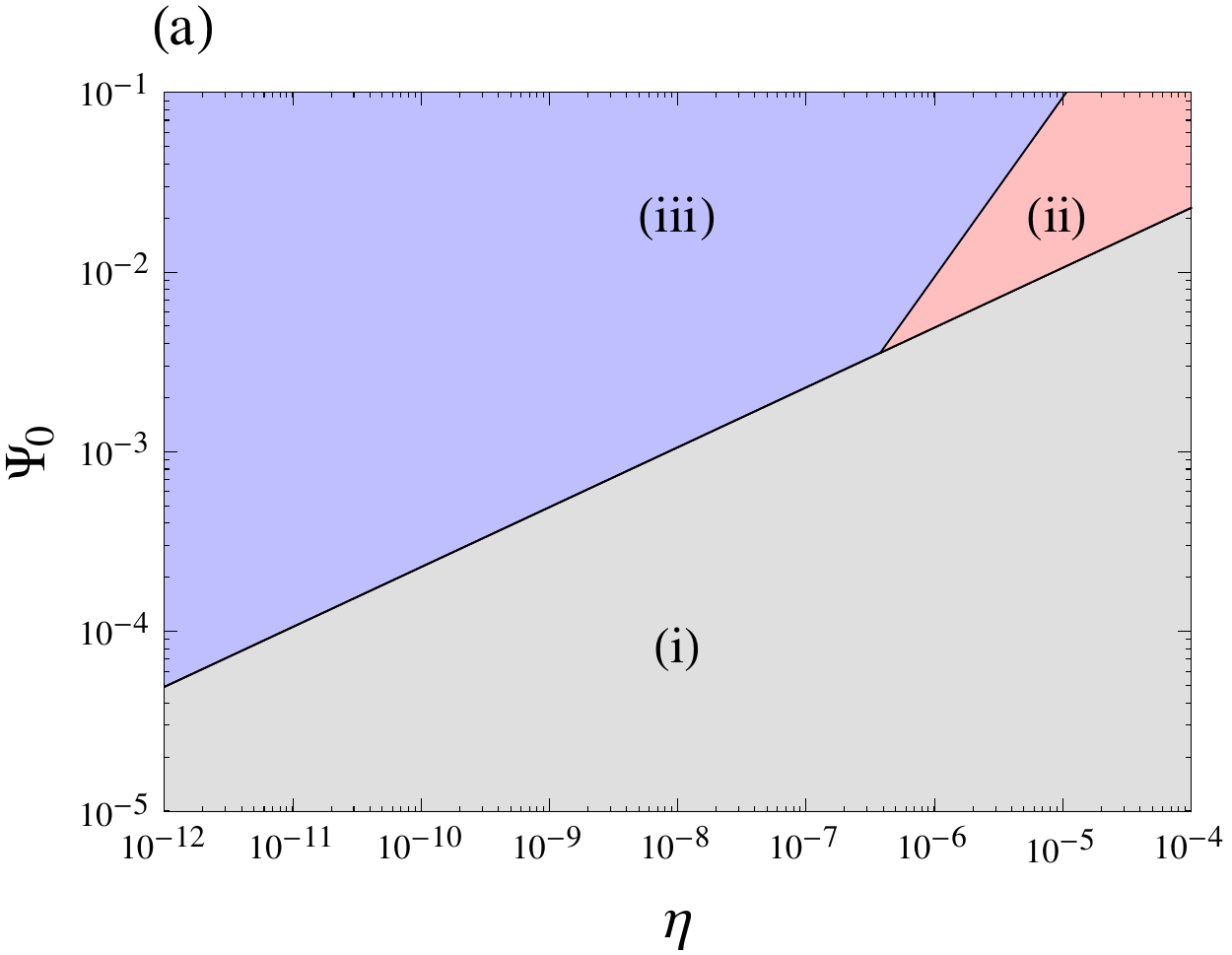}
\includegraphics[bb =1 4 357 278, width=8.2cm]{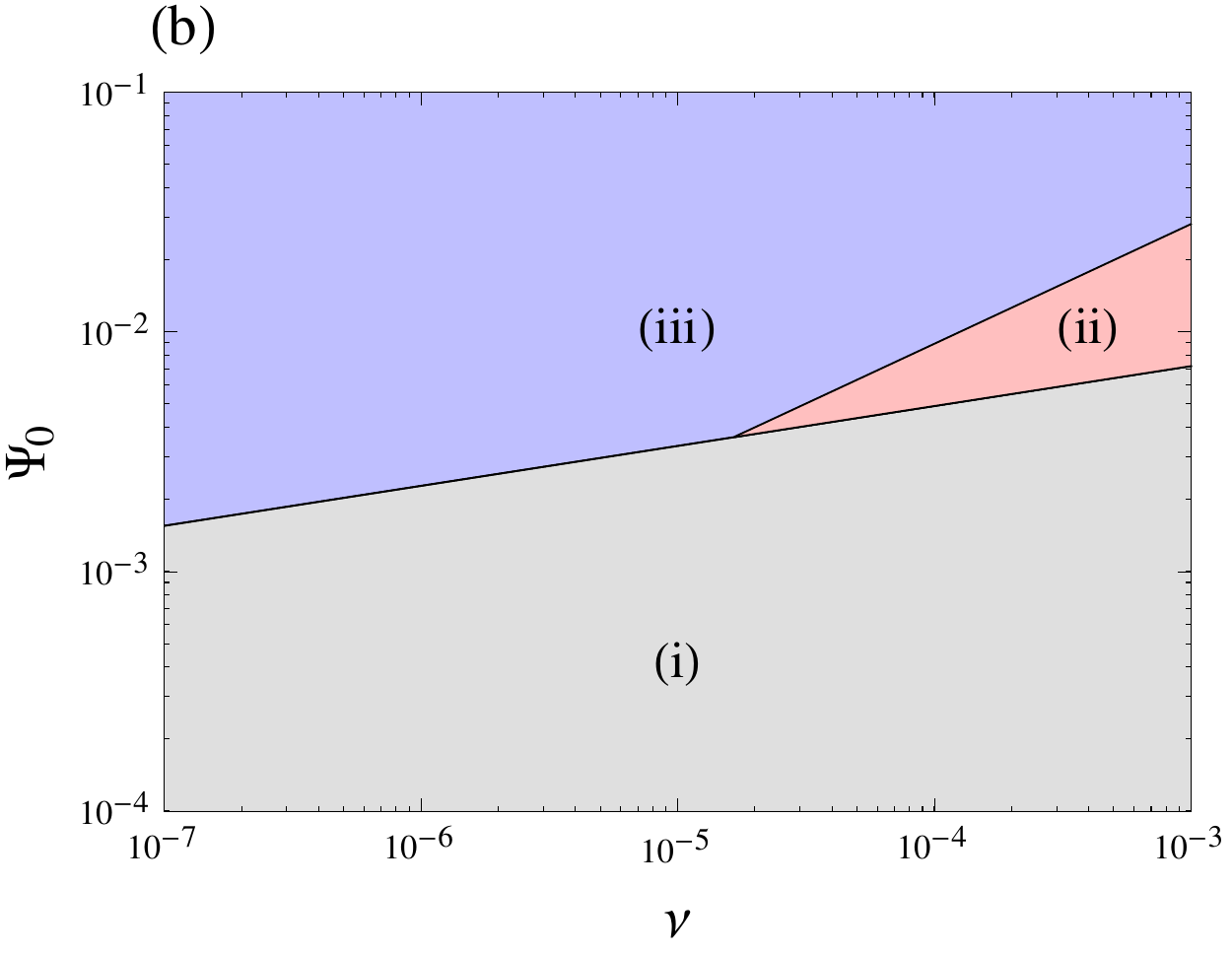}
\end{center}
\caption{(Color online) Two dimensional slices of a ``phase diagram'' for forced magnetic reconnection in the magnetohydrodynamical Taylor model. Fixed parameters are (a) $k=1/8$, $\nu/\eta=10$, and (b) $k=1/8$, $\eta=10^{-7}$. Three possible evolutions can occur: (i) Hahm-Kulsrud scenario \cite{HK_1985,Fitz_2003}, (ii) Wang-Bhattacharjee scenario \cite{WB_1992,Fitz_2003}, (iii) New scenario characterized by a plasmoid phase. These different scenarios of the forced reconnection process are summarized in Sec. \ref{sec6}.}
\label{fig2}
\end{figure}
In the following, we consider the results of simulations with $L_y = 16 \pi$ (i.e., $k=1/8$), $\tau=2$ and different values of $\Psi_0$, $\eta$, $\nu$. The choice of the parameters is such to fall within the three scenarios presented in the previous section. Indeed, for a fixed perturbation wave number $k$, the possible evolutions of the system can be organized in a three-dimensional ``phase diagram'' with $\Psi_0$, $\eta$ and $\nu$ on the three axis (a four-dimensional ``phase diagram'' is needed to retain also the $k$ dependence). Such a diagram can be constructed from the conditions derived in Sec. \ref{sec4} and summarized in Sec. \ref{sec6}. Figs. \ref{fig2}(a) and \ref{fig2}(b) show two slices of this diagram (the Hahm-Kulsrud scenario, which occurs if $\Psi_0 \ll \Psi_W$, is supposed to hold until $\Psi_0 = \Psi_W /3$), which can be used to graphically organize the knowledge of the reconnection dynamics in a system with given plasma and forcing parameters.

\begin{figure}
\begin{center}
\includegraphics[bb = 0 0 428 253, width=8.6cm]{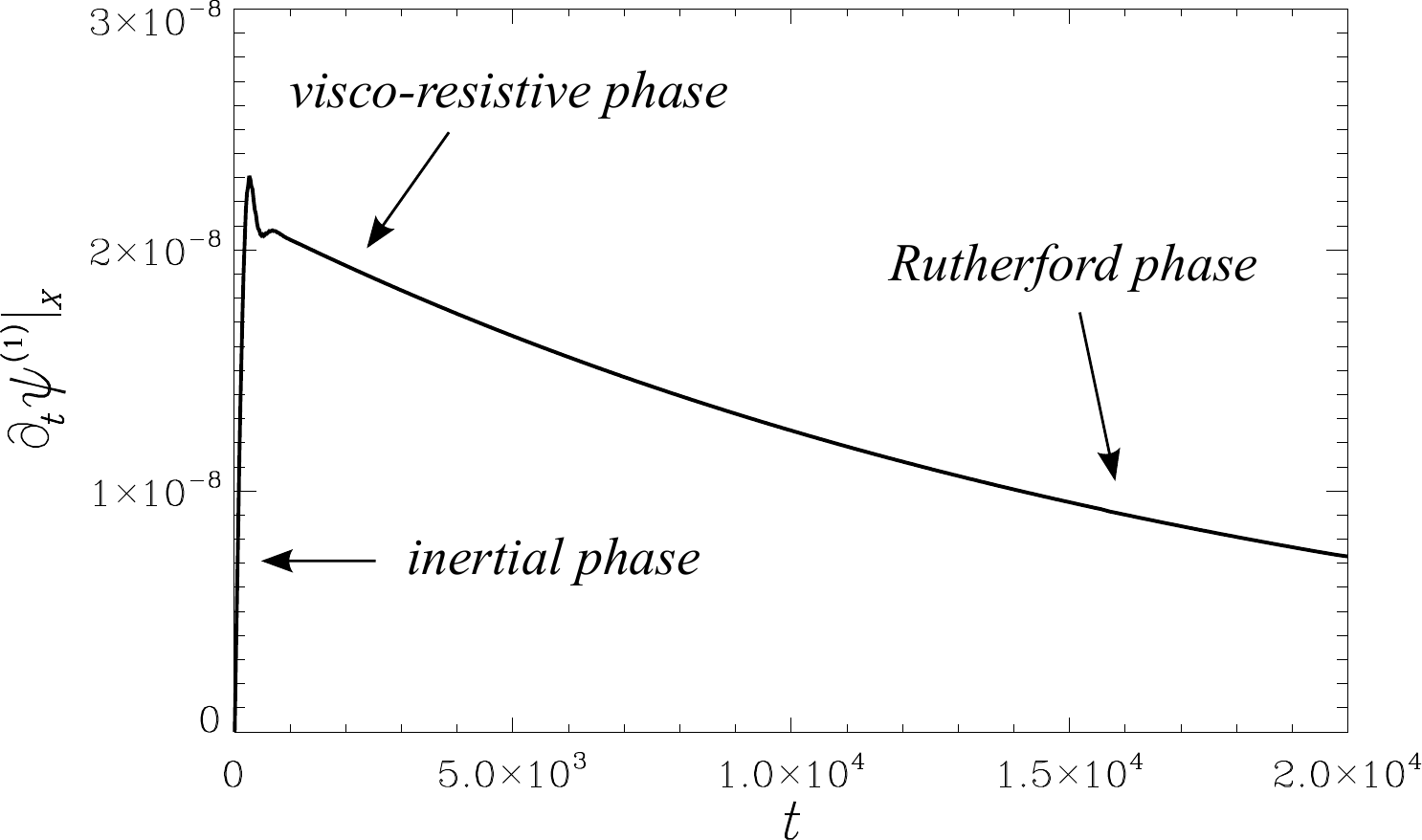}
\end{center}
\caption{Reconnection rate ${\partial_t}{\psi^{(1)}} |_X$ as a function of time from a numerical simulation with $\eta = 2 \times 10^{-6}$, $\nu = 2 \times 10^{-5}$, $\Psi_0= 4 \times 10^{-4}$.}
\label{fig3}
\end{figure}
\begin{figure}
\begin{center}
\includegraphics[bb = 0 0 420 253, width=8.6cm]{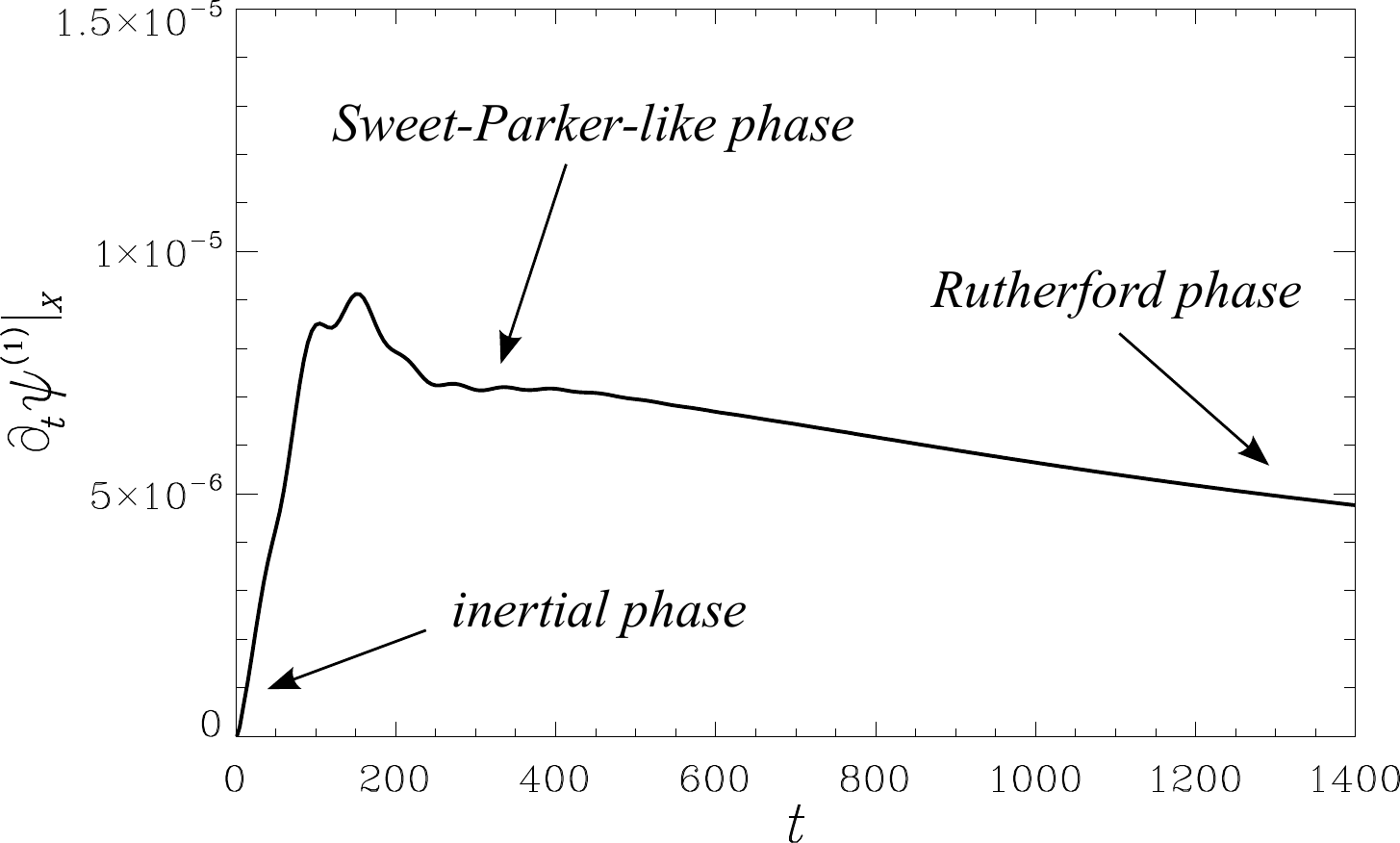}
\end{center}
\caption{Reconnection rate ${\partial_t}{\psi^{(1)}} |_X$ as a function of time from a numerical simulation with $\eta = 8 \times 10^{-6}$, $\nu = 8 \times 10^{-5}$, $\Psi_0= 4 \times 10^{-2}$.}
\label{fig4}
\end{figure}
\begin{figure}
\begin{center}
\includegraphics[bb = 0 0 416 252, width=8.6cm]{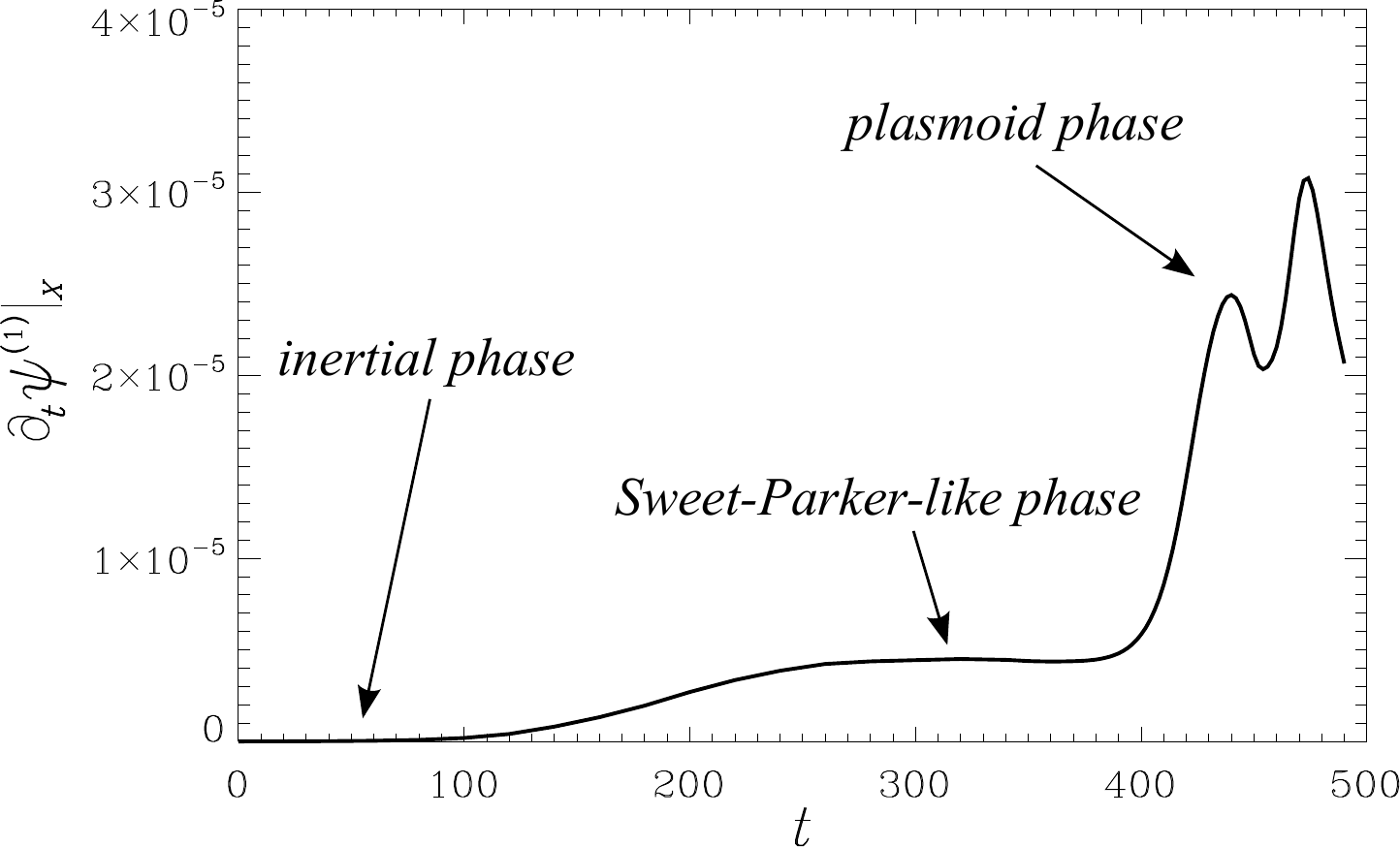}
\end{center}
\caption{Reconnection rate ${\partial_t}{\psi^{(1)}} |_X$ as a function of time from a numerical simulation with $\eta = 4 \times 10^{-8}$, $\nu = 4 \times 10^{-7}$, $\Psi_0= 4 \times 10^{-2}$.}
\label{fig5}
\end{figure}
The typical evolution of the reconnection rate when the perturbation amplitude satisfies condition (\ref{Rutherford_Condition_2}) (but is not so small as to prevent the system from entering into the nonlinear regime) is shown in Fig. \ref{fig3}. In this case, a current sheet builds up at the resonant surface during the inertial phase, which lasts until $t < \tau_i \approx 147$. Then, after the magnetic island attains a constant-$\psi$ behavior at $t \sim \tau_{vr} \approx 684$, the reconnection process evolves through a visco-resistive phase as long as the island width becomes comparable to the linear layer width at $t \sim 7000$. After this time the system proceeds nonlinearly according to a slow Rutherford evolution as predicted by Hahm and Kulsrud \cite{HK_1985}.

A different evolution is shown in Fig. \ref{fig4}. In this case, the perturbation amplitude satisfies condition (\ref{SP_Condition_2}) but not condition (\ref{Plasmoid_Condition_3}). After the initial inertial phase ($t < \tau_i \approx 93$), the system passes into the nonlinear regime at $t \sim 90$ while the magnetic island is still characterized by a non-constant-$\psi$ behaviour. As a consequence, the reconnection process evolves according to a Sweet-Parker-like regime and only subsequently a Rutherford regime takes place, as predicted by Wang and Bhattacharjee \cite{WB_1992}

\begin{figure}
\begin{center}
\includegraphics[bb =225 210 363 739, width=4.2cm]{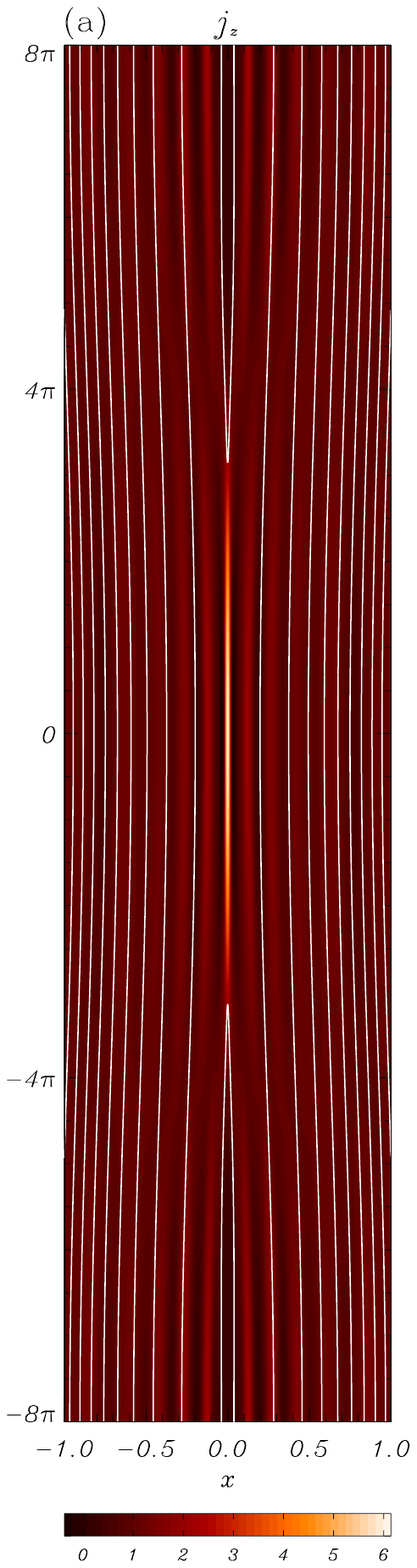}
\includegraphics[bb =225 210 363 739, width=4.2cm]{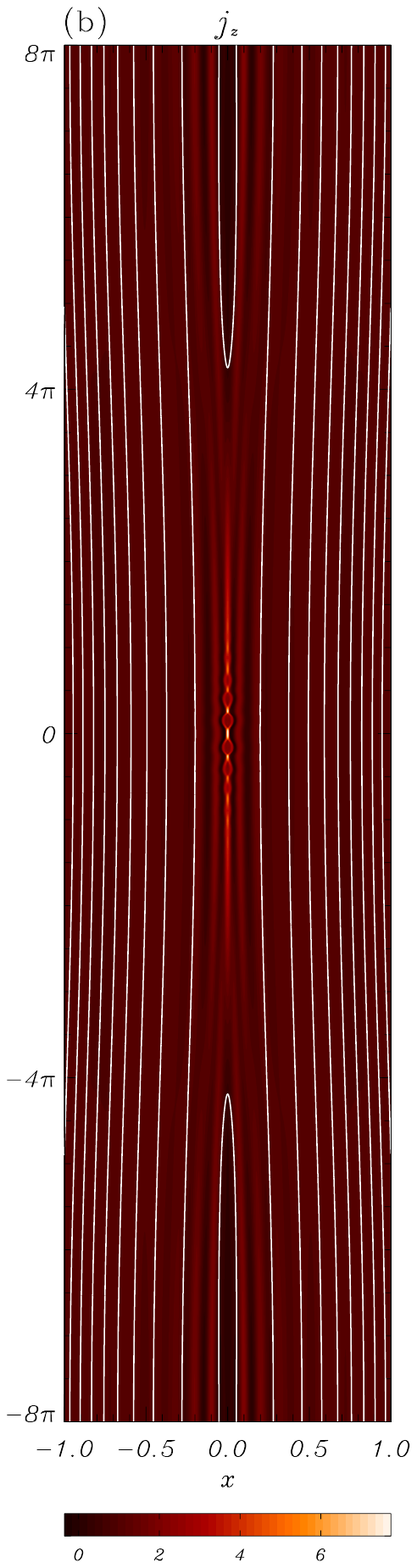}
\end{center}
\caption{(Color online) From the numerical simulation shown in Fig. \ref{fig5}, contour plots of the out-of-plane current density $j_z (x,y)$ with the in-plane component of some magnetic field lines (white lines) superimposed at (a) $t=320$ and (b) $t=450$.}
\label{fig6}
\end{figure}
\begin{figure}
\begin{center}
\includegraphics[bb = 0 0 396 256, width=8.6cm]{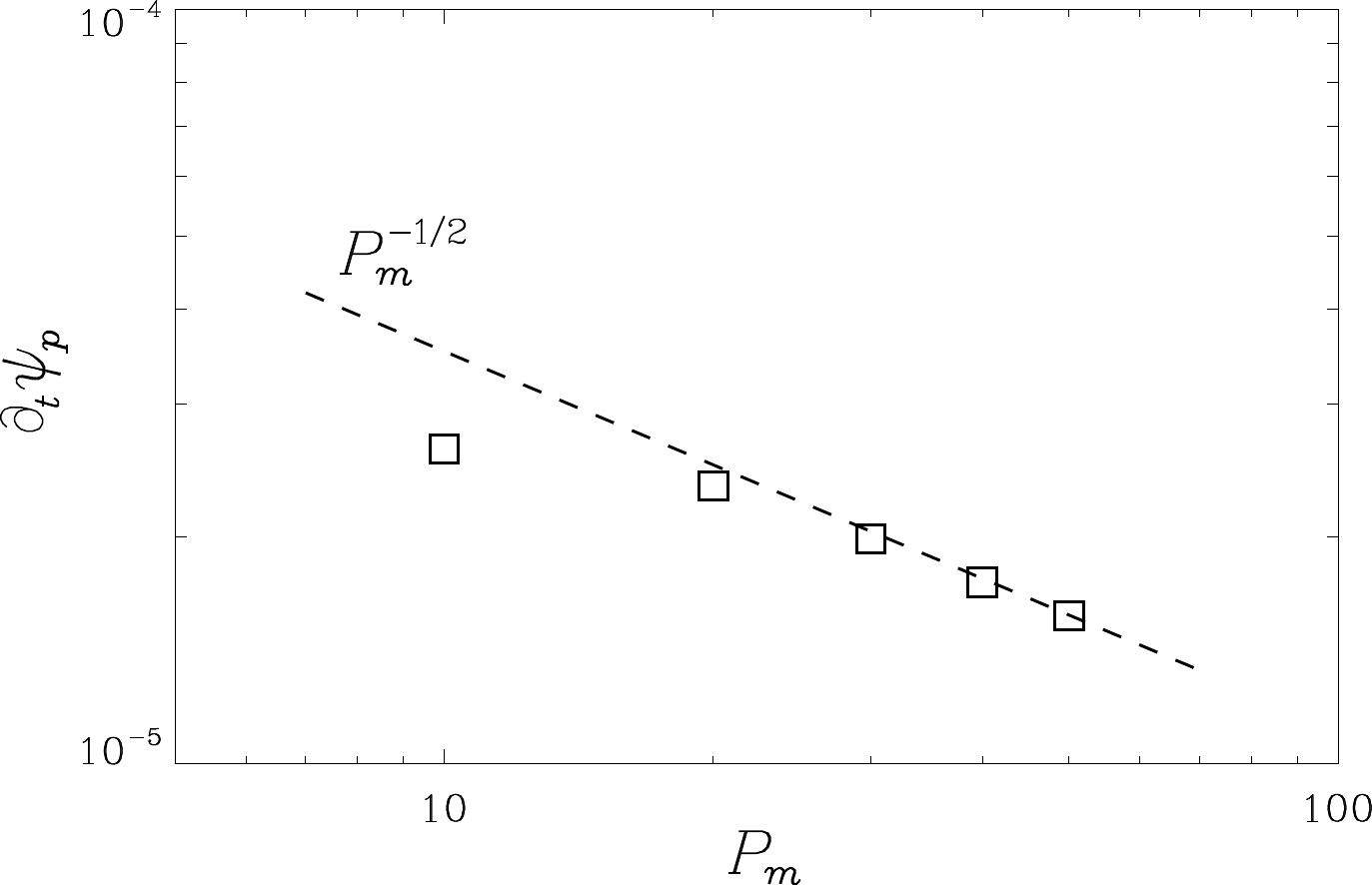}
\end{center}
\caption{Time-averaged reconnection rate in the plasmoid-dominated regime ${\partial_t}{\psi_p}$ as a function of the magnetic Prandtl number $P_m$ for $\eta = 4 \times 10^{-8}$. The theoretical slope $P_m^{-1/2}$ is shown for reference.}
\label{fig7}
\end{figure}
A completely different scenario occurs when the perturbation amplitude satisfies condition (\ref{Plasmoid_Condition_4}), as shown in Fig. \ref{fig5}. In this case the inertial phase lasts until the non-constant-$\psi$ island width becomes comparable to the linear layer width, at $t \sim 150$ (well before $\tau_i \approx 543$). Then the system proceeds into the nonlinear regime according to a Sweet-Parker-like evolution characterized by an elongated current sheet with slowly varying aspect ratio. The elongated current sheet that characterizes this stage of the reconnection process is shown in Fig. \ref{fig6}(a). However, as the reconnecting current sheet gradually thins, the plasmoid instability sets in at $t \approx 360$. A sudden increase of the reconnection rate occurs when the first nonlinear plasmoids reach the width of the current sheet, which begins to break up at $t \approx 410$. Then, after $t \approx 440$, when the plasmoid chain is well developed (see Fig. \ref{fig6}(b)), the reconnection rate seems to fluctuate around a constant mean value, as expected on the basis of previous numerical simulations of plasmoid-dominated reconnection \cite{BHYR_2009,Cassak_2009,HB_2010,HBS_2011,Loureiro_2012,Murphy_2013}. The growth of the plasmoids is so fast that they start to merge before being ejected in the downstream region, and the simulation is stopped at $t = 490$, when the adopted truncated Fourier expansion can no longer resolve the large gradients in the $y$ direction that are due to the coalescence of the plasmoids in this stage of the reconnection process.

Although in our simulations we are not able to achieve a long statistical steady-state of the plasmoid phase, it is interesting to try to look at the dependence on the magnetic Prandtl number of the reconnection rate in this phase. Performing numerical simulations by varying $P_m$, we obtain that the time-averaged reconnection rate in the plasmoid phase scales roughly as $P_m^{-1/2}$ when the magnetic Prandtl number is large (see Fig. \ref{fig7}), in agreement with the theoretical scaling (\ref{Rec_rate_plasmoids_at_X_3}) in the limit $P_m \gg 1$. The reconnection rate for the simulation with $P_m = 10$ deviates more from the theoretical predictions mainly because of the numerical dissipation that adds up to the physical dissipation and that is more significant when the physical viscosity is lower.

\section{Summary} \label{sec6}
In this work we have examined a fundamental problem of forced magnetic reconnection, the so called Taylor problem, in which magnetic reconnection is driven by a small amplitude boundary perturbation in a tearing-stable slab plasma equilibrium. In particular, we have shown how the evolution of the reconnection process depends on both the external source perturbation and the microscopic plasma parameters. 

Three possible evolutions can occur in a magnetohydrodynamical plasma.

i) Hahm-Kulsrud scenario \cite{HK_1985,Fitz_2003}: if the perturbation is such that $\Psi_0 \ll \left( {k^{-1/3}/{\Delta '_s}} \right)   {\left( {\nu \eta } \right)^{1/6}}$, the reconnection evolves initially through a linear inertial phase, then through a linear visco-resistive phase, and finally (if the perturbation is not excessively small) it proceeds in the nonlinear regime according to a Rutherford evolution.

ii) Wang-Bhattacharjee scenario \cite{WB_1992,Fitz_2003}: if the perturbation is such that $({\Delta '_s}{k^{1/3}})^{-1} {\left( {\nu \eta } \right)^{1/6}}  \lesssim  \Psi_0 \leq C \left( {k/{\Delta '_s}} \right) \eta {\left( 1 + {\nu/ \eta } \right)^{1/2}}$, after the linear inertial phase the system passes into the nonlinear regime according to a Sweet-Parker-like evolution, which then gives way to the Rutherford regime.

iii) Our scenario: if the perturbation is such that $\Psi_0 \gtrsim ({\Delta '_s}{k^{1/3}})^{-1}  {\left( {\nu \eta } \right)^{1/6}}$ and $\Psi_0 > C \left( {k/{\Delta '_s}} \right) \eta {\left( 1 + {\nu/ \eta } \right)^{1/2}}$, after the linear inertial phase and an initial nonlinear evolution with a gradually evolving current sheet, the reconnection suddenly enters into a fast plasmoid phase. Note that the realization of a Sweet-Parker sheet before the onset of the plasmoid phase is not mandatory, since in systems with extremely large Lundquist numbers the plasmoid instability may be triggered prior to the attainment of such current sheet \cite{Pucci_2014,Tenerani_2015,Uzdensky_2014}

The plasmoid formation plays a key role in allowing fast magnetic reconnection by making the reconnection rate independent of the Lundquist number, which is known to be extremely large in many astrophysical and laboratory plasmas. While for the case of negligible viscosity the reconnection rate is also independent of the magnetic Prandtl number, we have shown that when viscosity is not negligible  ${\partial _t}{\psi _p} \approx {\epsilon _c}{\left( {{\Delta '_s}{\Psi _0}} \right)^2} {\left( {1 + {\nu}/{\eta}} \right)^{-1/2}}$. Therefore, the reconnection rate in the visco-resistive plasmoid phase decreases as the magnetic Prandtl number increases. The reconnection rate depends also on the perturbation amplitude, whose effect is to increase the magnetic field upstream of the reconnection region.

It is important to highlight that there exists a critical perturbation wave number $k^*$ below which the evolution of the system always leads to the plasmoid phase. Since $k^*$ increases for decreasing values of the plasma resistivity and viscosity, the development of stable elongated current sheets is not attainable in many situations of interest in astrophysical and laboratory fusion plasmas. Therefore, also modest perturbation amplitudes may lead to plasmoid-dominated reconnection in plasmas characterized by very small resistivity and viscosity.

\acknowledgments

The authors would like to acknowledge fruitful conversations with Dario Borgogno, Daniele Brunetti, Luke Easy, Richard Fitzpatrick, Enzo Lazzaro and Fulvio Militello. One of us (L.C.) is grateful for the hospitality of the Institute for Fusion Studies at the University of Texas at Austin, where part of this work was conceived.
This work was carried out under the Contract of Association Euratom-ENEA and was also supported by the U.S. Department of Energy under Contract No. DE-FG02-04ER-54742.

\end{document}